\documentclass[aps,prl,reprint,amsmath,amssymb,superscriptaddress,longbibliography,floatfix,twocolumn]{revtex4-2}

\usepackage{graphicx}
\usepackage{mathtools}
\usepackage{float}
\usepackage[dvipsnames]{xcolor}
\usepackage[bookmarks=true,colorlinks,linkcolor=OrangeRed,urlcolor=NavyBlue,citecolor=NavyBlue]{hyperref}
\usepackage{orcidlink}
\usepackage{braket}
\usepackage{printlen}

\newcommand{\defaulttensorsize}{10pt}
\newcommand{\tensorsize}{10pt}

\usetikzlibrary{shapes.geometric}
\usetikzlibrary{shapes.misc}
\tikzstyle{tensor}=[draw, inner sep=0, outer sep=0, minimum size=\tensorsize]
\tikzstyle{notensor}=[inner sep=0, outer xsep=2pt, outer ysep=0, minimum size=\tensorsize]
\tikzstyle{atensor}=[tensor, circle]
\tikzstyle{ctensor}=[tensor, circle]
\tikzstyle{dtensor}=[tensor, diamond]
\tikzstyle{wtensor}=[tensor]
\tikzstyle{ltensor}=[tensor, rounded rectangle, rounded rectangle left arc=none]
\tikzstyle{rtensor}=[tensor, rounded rectangle, rounded rectangle right arc=none]
\tikzstyle{etensor}=[tensor, minimum height=(1cm/\defaulttensorsize*0.5*2+1)*\tensorsize]
\tikzstyle{widetensor}[2]=[tensor, minimum width=(1cm/\defaulttensorsize*0.75*(#1-1)+1)*\tensorsize, xshift=(1cm/\defaulttensorsize*0.75*(#1-1)/2*\tensorsize]

\tikzstyle{tensornetwork}=[baseline=-0.25em, xscale=0.75, yscale=0.5, scale=1,
                           every node/.style={inner sep=0, outer xsep=2pt, outer ysep=5pt, text depth=0pt},
                           execute at end picture={\path (current bounding box.south west) +(-2pt, 0) (current bounding box.north east) +(2pt, 0);}]

\renewcommand{\refeq}[1]{Eq.~(\ref{#1})}
\newcommand{\reffig}[1]{Fig.~[\ref{#1}]}

\newcommand{\kept}{{\scriptstyle \mathrm{K}}}
\newcommand{\disc}{{\scriptstyle \mathrm{D}}}

\begin{document}

\title{Comment on ``Controlled Bond Expansion for Density Matrix Renormalization Group Ground State Search at Single-Site Costs'' (Extended Version)}

\author{Ian P.~McCulloch${}^{\orcidlink{0000-0002-8983-6327}}$}
\email{ian@phys.nthu.edu.tw}
\affiliation{Department of Physics, National Tsing Hua University, Hsinchu 30013, Taiwan}

\author{Jesse J.~Osborne${}^{\orcidlink{0000-0003-0415-0690}}$}
\email{j.osborne@uqconnect.edu.au}
\affiliation{School of Mathematics and Physics, The University of Queensland, St.~Lucia, QLD 4072, Australia}

\maketitle

In a recent Letter\cite{CBE}, Gleis, Li, and von Delft present an algorithm for expanding the bond dimension of a Matrix Product State wave function, giving accuracy similar to 2-site DMRG, but computationally more efficient, closer to the performance of 1-site DMRG. The Controlled Bond Expansion (CBE) algorithm uses the Hamiltonian projected onto two sites (which we refer to as the \emph{environment site} and the \emph{active site}), $H^{2\text{s}}\psi^{2\text{s}}$, and then further projected onto the two-site tangent space, to extract a set of $k$ vectors that are used to expand the basis between the two sites\footnote{We use $k$ rather than $\tilde{D}$ for clarity of notation}. CBE achieves this with a complicated sequence of five singular value decompositions (SVDs), in order to project onto the 2-site tangent space and reduce the bond dimension of the tensor network such that the contraction can be done in time $O(dwD^3)$, where $w$ is the MPO bond dimension, $d$ is the local Hilbert space size, and $D$ is the MPS bond dimension. This is in addition to the usual SVD for the DMRG truncation procedure. In this Comment, we show that (1) the projection onto the 2-site tangent space is unnecessary, and is generally not helpful; (2) the sequence of 5 SVDs can be replaced by a single $QR$ decomposition (optionally with one SVD as well), making use of the randomized SVD (RSVD)\cite{RSVD} with high accuracy and significantly improved efficiency, scaling as $O(dwkD^2)$
i.e., the most expensive operations are only quadratic in the bond dimension $D$ and linear in the number of expansion vectors $k$; (3) several statements in Ref.~\cite{CBE} about the variational properties of the CBE algorithm are incorrect, and the variational properties are essentially identical to existing algorithms including 2-site DMRG and single-site subspace expansion (3S)\cite{3S}; (4) a similar RSVD approach can be applied to the 3S algorithm, which leads to many advantages over CBE, especially in systems with long range interactions. We also make some comments on the benchmarking MPS algorithms, and the overall computational efficiency with respect to the accuracy of the calculation.

To briefly review the CBE approach, consider one step of DMRG on a right-to-left sweep. We can express the wavefunction as,
\begin{equation}
     \ket{\Psi} =
    \begin{tikzpicture}[tensornetwork,baseline=-0.25em]
        \node[notensor]                       (B) at (0, 0) {\(\ldots\)};
        \node[ltensor, label=below:\(A\)]     (L) at (1, 0) {};
        \node[ctensor, label=below:\(C\)]     (C) at (2, 0) {};
        \node[notensor]                       (E) at (3, 0) {\(\ldots\)};
        \draw (L.north) -- +(0, 0.25);
        \draw (C.north) -- +(0, 0.25);
        \draw (B) -- (L);
        \draw [ultra thick] (L) -- (C);
        \draw (C) -- (E);
    \end{tikzpicture}
    \; ,
   \label{eq:psi}
\end{equation}
where the previous step optimized the site to the right of $C$ and shifted the orthogonality center to site $C$.
At the current step, site $C$ is the active site and $A$ is the left-orthogonal environment site. As a starting point to expanding the bond dimension between $A$ and $C$, we can consider the action of the Hamiltonian on these two sites (effectively, a single iteration of 2-site DMRG)\cite{Valentin}:
\begin{equation}
    \begin{tikzpicture}[tensornetwork]
        \node[etensor, rounded corners] (E) at (-1, 0) {};
        \node[notensor] (Lconj) at (0, 1) {};
        \node[ltensor, label=below:\(A\)]     (L) at (0,-1) {};
        \node[ctensor, label=below:\(C\)]     (C) at (1,-1) {};
        \node[wtensor] (H1) at (0, 0) {};
        \node[wtensor] (H2) at (1, 0) {};
        \node[notensor] (Cconj) at (1, 1) {};
        \node[etensor, rounded corners] (F) at (2, 0) {};
        \draw (L) -- (H1) -- (Lconj);
        \draw (C) -- (H2) -- (Cconj);
        \draw (E) -- (H1) -- (H2) -- (F);
        \draw (E.east |- L) -- (L) -- (C) -- (F.west |- C);
        \draw (E.east |- Lconj) -- +(0.25,0);
        \draw (F.west |- Cconj) -- +(-0.25,0);
    \end{tikzpicture}
        =
    \begin{tikzpicture}[tensornetwork]
        \node[widetensor=2, rounded corners, label=below:\(X\)] (C) at (0,0) {};
        \coordinate (Lconj) at (0, 1) {};
        \coordinate (Cconj) at (1, 1) {};
        \coordinate (B) at (-1, 0) {};
        \coordinate (E) at (3, 0) {};
        \draw (C.north -| Lconj) -- (Lconj);
        \draw (C.north -| Cconj) -- (Cconj);
        \draw (C.east) -- +(0.25,0);
        \draw (C.west) -- +(-0.25,0);
    \end{tikzpicture}
    \; ,
    \label{eq:MatVec}
\end{equation}
where the final tensor is denoted $X$, which we treat as a matrix reshaped to be have dimensions $dD \times dD$. We could do an SVD of $X$ and construct the $D+k$ most important left singular vectors, and use these to construct a new left-orthogonalized tensor $A'$, thereby increasing the bond dimension from $D$ to $D+k$. However this is very expensive, $O(d^3D^3)$, plus the cost of constructing $X$, so we look for a way to reduce the computational complexity. Firstly, it will generally be the case that the leading left singular vectors of $X$ have a large overlap with the states already in $A$, so we could keep those vectors and add to them the $k$ largest singular values of $X$ that are orthogonal. To do that, we project out the states in $A$ via $X_{\bar{A}} = (I - A A^\dagger)X = X - A (A^\dagger X)$, treating $A$ as a $dD \times D$ matrix ($X_{\bar{A}}$
is referred to as $C^{\text{tmp}}_{l+1}$ in Ref.~\cite{CBE} Algorithm 1.) Once we have the $k$ largest singular vectors of $X_{\bar{A}}$, we add these as new columns of $A$, increasing its dimensions to $dD \times (D+k)$,
and add corresponding environment Hamiltonian $E$-matrix elements at a cost of $O(dwkD^2)$. The next problem is how to contract the tensor network to obtain $X_{\bar{A}}$ faster than $O(d^2 w D^3)$. CBE achieves this by a sequence of three SVDs to obtain an isometry $S$ (referred to as \emph{shrewd selection} in Ref.~\cite{CBE}) that projects the left hand side of $X_{\bar{A}}$ from $dD$ states to $\hat{D} \simeq D$,
\begin{equation}
   SX_{\bar{A}}    =
    \begin{tikzpicture}[tensornetwork]
        \node[etensor, rounded corners] (E) at (-1, 0) {};
        \node[ltensor, label=\(S\)] (Lconj) at (0, 1) {};
        \node at (0.65,1+0.07) {\(\hat{D}\)};
        \node[ltensor]     (L) at (0,-1) {};
        \node[ctensor]     (C) at (1,-1) {};
        \node[wtensor] (H1) at (0, 0) {};
        \node[wtensor] (H2) at (1, 0) {};
        \node[notensor] (Cconj) at (1, 1) {};
        \node[etensor, rounded corners] (F) at (2, 0) {};
        \draw (L) -- (H1) -- (Lconj);
        \draw (C) -- (H2) -- (Cconj);
        \draw (E) -- (H1) -- (H2) -- (F);
        \draw (E.east |- L) -- (L) -- (C) -- (F.west |- C);
        \draw (E.east |- Lconj) -- (Lconj);
        \draw (Lconj.east) -- +(0.25,0);
        \draw (F.west |- Cconj) -- +(-0.25,0);
    \end{tikzpicture}
    \; .
\end{equation}
At this point the contraction can be done in cost $O(dwD^3)$, which is the same scaling as an iteration of single-site DMRG.


\noindent \textbf{2-site tangent space --}
The CBE algorithm includes a second projection, to project out the states used in the $C$ tensor from $X_{\bar{A}}$ prior to the SVD. This can be expressed as another projection $X_{\bar{A}}(I - V V^\dagger)$,  where $V^\dagger$ is the matrix of right singular vectors of $C$ ($V^\dagger$ is referred to as $X^{\text{orth}}_{l+1}$ in Ref.~\cite{CBE} Algorithm 1). This requires another SVD to obtain $V^\dagger$.  This projection appears to be of little benefit, in fact it is not hard to construct examples where this could cause a problem. For example, consider an iteration of DMRG starting from an initial vacuum state $\ket{\cdots 0 0 \cdots}$, and suppose on the two sites corresponding to $A$ and $C$ the Hamiltonian contains some matrix element $\ket{10}\bra{00}$.
In expanding the bond dimension of the $A$ site, the state $\ket{1}$ is clearly a good choice, and the state $\ket{10}$ will be represented in the $X_{\bar{A}}$ tensor. However the CBE algorithm projects out the kept states on the \emph{right} site as well, meaning any state of the form $\ket{x0}$ will be projected out, since $\ket{0}$ will be one of the right singular vectors of $C$. Hence the state $\ket{10}$ that should be included in the expanded basis will be missing. Important states that would appear in a 2-site DMRG update are therefore excluded. This is not a contrived example, as similar constructions appears in many constrained models such as gauge theories and quantum link models, typically with terms that span 3 sites (e.g., creation of a particle-antiparticle pair with an accompanying change in flux at a gauge site). In some cases the missing basis vectors would be included with additional DMRG sweeps, but this is less efficient and a real problem with TDVP\cite{TDVPOriginal} where such additional sweeps are not usually performed\cite{TDVPLubich}.
This problem can be avoided by simply omitting the projection onto the null space of $C$, and incidentally saving several $O(dD^3)$ operations. We have not found any numerical examples where omitting this projection has a detrimental effect on the calculation.

Throughout the Letter, the authors refer to selecting expansion vectors from the orthogonal two-site tangent space, referred to as $\disc\disc$, the space of states that are discarded both on the $A$ site and the $C$ site\footnote{
Abstract: CBE identifies parts of the orthogonal space carrying significant weight in $H\Psi$ and expands bonds to include only these.
Page 2, final paragraph:  The first insight is that the subspace of DD relevant for lowering the GS energy is relatively small [...] When expanding a bond, it thus suffices to add only this small subspace (hence the moniker controlled bond expansion), or only part of it, to be called relevant $\disc\disc$ ($r_{\disc\disc}$). Also the following discussion onto page 3.
Footnote [29]:  If 2s DMRG has converged to an optimal MPS $\Psi_\disc$ with fixed bond dimension $D$, the size of $r_{\disc\disc}$ is zero.}.
However the left singular vectors of $X_{\bar{A}}$ cannot be separated into components that are kept (or not) with respect to site $C$. In order to split the tensor $X$ into 2-site spaces $\kept\kept$, $\kept\disc$, $\disc\kept$, $\disc\disc$, we need to first shift the orthogonality center onto the bond, by factorizing $C = \Lambda B$, where $B$ is right orthogonalized and $\Lambda$ lives on the bond between $A$ and $B$.
%
Extending this bond basis with the null spaces $\bar{A}$ and $\bar{B}$ gives the 2-site tangent spaces as the four different combinations of $(A \oplus \bar{A}) \otimes (B \oplus \bar{B})$. Expansion vectors from $\bar{A}$ cannot be selected by whether they belong (or not) to $\bar{B}$, since they are separate Hilbert spaces and that information is lost when taking the tensor product. The Hilbert space that will be optimized in the DMRG iterations is the $(D+k)\times dD$-dimensional space of
$(A \oplus \tilde{A}) \otimes (B \oplus \bar{B})$, where $\tilde{A} \subseteq \bar{A}$
is the carrier space of the $k$ expansion vectors.
Writing the expansion tensor $X$ in partitioned form,
\begin{equation}
   X =
   \begin{pmatrix} X_{\kept\kept} & X_{\kept\disc} \\ X_{\disc\kept} & X_{\disc\disc} \end{pmatrix}
   \; ,
\end{equation}
the states already contained in $A$ are the top rows of $X$, $\begin{pmatrix} X_{\kept\kept} & X_{\kept\disc} \end{pmatrix}$. Candidate expansion vectors come from the bottom rows, being the null space $\bar{A}$. This corresponds to the SVD of the $(d-1)D \times dD$ matrix $\begin{pmatrix} X_{\disc\kept} & X_{\disc\disc} \end{pmatrix}$.
CBE chooses to only consider the component $X_{\disc\disc}$ by projecting onto the discarded states $\bar{B}$, which sets $X_{\disc\kept}$ to zero prior to the SVD. This throws away potentially useful information that could be used to improve the basis, and does not have the effect that the authors suggest.

\noindent \textbf{Randomized SVD --} Instead of using shrewd selection\cite{CBE} to approximate the contraction of \refeq{eq:MatVec}, a more efficient and accurate way to obtain the $k$ dominant singular values of $X_{\bar{A}}$ is via the Randomized SVD (RSVD)\cite{RSVD}. the RSVD has already been used to accelerate 2-site DMRG\cite{Tree_RSVD}, and other algorithms\cite{TEBD_RSVD,TNR_RSVD,TT_RSVD}. The core of the RSVD is the \emph{range finding algorithm} for finding an approximate span of the dominant singular vectors of a matrix.
To find the span of the dominant $k$ left singular vectors of $X_{\bar{A}}$, we multiply it by a Gaussian random matrix $\Omega$ of dimension $dD \times k$. We then perform a $QR$ decomposition,
\begin{equation}
    QR = X_{\bar{A}} \Omega \; ,
\end{equation}
where $Q$ is a $dD \times k$ left-orthogonal matrix and $R$ is $k \times k$ upper triangular. $Q$ contains an approximation of the dominant left singular vectors of $X_{\bar{A}}$. If we only need $k$ \emph{approximate} singular vectors then we can stop here. But while the leading singular vectors should be represented quite accurately, the tail of the spectrum will not be accurate. To obtain an accurate spectrum for all $k$ singular values, the RSVD algorithm uses the range finding algorithm to firstly find a larger set of vectors $k+p$, where $p$ is the \emph{oversampling parameter} ($p=10$ is typical)\cite{RSVD}. Once we have $Q$ as a $dD \times (k+p)$ dimensional matrix, we project the original matrix $X_{\bar{A}}$ into this space and perform an SVD,
\begin{equation}
    U D V^\dagger = Q^\dagger X_{\bar{A}} \; ,
\end{equation}
where $D$ is the diagonal matrix of singular values, $U$ is left-orthogonal, and $V^\dagger$ is right-orthogonal. Projecting $U$ onto the $k$ largest singular values, $QU$ now contains the $k$ dominant singular vectors of $X_{\bar{A}}$ with high accuracy. However we have found that this step is often not necessary, and in our numerical experiments the range finding algorithm works quite well and is very cheap as it requires just one tensor network contraction and one $QR$ decomposition. The reason that the range finding works so well is that the selection of states to use as expansion vectors is not very sensitive, and in fact \emph{random} vectors work reasonably well, as shown below, and also used to good effect by Oseledets and Dolgov\cite{RandEnrichment}.
Range finding can be seen as a semi-random algorithm that improves upon purely random vectors. RSVD interpolates between range finding for $p \rightarrow 0$, becoming equivalent to the exact SVD when $k+p$ is equal to the rank of $X_{\bar{A}}$. Importantly, the computational complexity of the RSVD and range finding algorithms are dominated by the matrix-matrix multiply of $\Omega$, which only has $k$ columns. Inserting $\Omega$ into the tensor network results in a contraction that has a computational complexity of $O(dwkD^2) + O(d^2 w^2 kD)$,
\begin{equation}
    X \Omega =
        \begin{tikzpicture}[tensornetwork]
        \node[etensor, rounded corners] (E) at (-1, 0) {};
        \node[ctensor, label=\(\Omega\)] (Omega) at (1, 1) {};
        \node[ltensor]     (L) at (0,-1) {};
        \node[ctensor]     (C) at (1,-1) {};
        \node[wtensor] (H1) at (0, 0) {};
        \node[wtensor] (H2) at (1, 0) {};
        \node[notensor] (Lconj) at (0, 1) {};
        \node[etensor, rounded corners] (F) at (2, 0) {};
        \draw (L) -- (H1) -- (Lconj);
        \draw (C) -- (H2) -- (Cconj);
        \draw (E) -- (H1) -- (H2) -- (F);
        \draw (E.east |- L) -- (L) -- (C) -- (F.west |- C);
        \draw (E.east |- Lconj) -- +(0.25,0);
        \draw (Omega.west) -- +(-0.25,0);
        \node at (0.3,1) {\(k\)};
        \draw (F.west |- Omega) -- (Omega);
    \end{tikzpicture}
    \; .
\end{equation}
Projecting $X\Omega$ onto the null space of $A$ can now be done at a cost of $O(dkD^2)$. So finally the most expensive operation is $O(dwkD^2)$\footnote{The projection to remove the right singular vectors could also be done with the same cost, however the SVD to obtain these singular vectors is relatively expensive, at a cost of $O(dD^3)$.}. If we choose $k$ to scale with $D$, e.g.\ $k = \delta D$, then formally the scaling is still $O(D^3)$, but in practice the prefactor $\delta$ is usually much smaller than 1 (the authors use $\delta=0.1$, which we also find works quite well), so the cost is negligible compared with the other stages of the DMRG algorithm.

\noindent \textbf{Variational properties --}
There are several statements in \cite{CBE} about the variational properties of the CBE algorithm that are incorrect, and the variational properties are largely the same as existing algorithms including 2-site DMRG and single-site subspace expansion (3S)\cite{3S}.
The authors state (bottom of page 3), ``The energy minimization based on $H^{1\text{s,ex}}_{l+1}$ is variational, hence each CBE update strictly lowers the GS energy''. This is a non sequitur. The sentence immediately preceding, CBE update step (iv), is a truncation of the bond dimension from $D + \tilde{D}$ back to $D$, which involves a loss of fidelity of the wavefunction and hence an inevitable increase in the energy\cite{Variational}. Similar incorrect claims are repeated in the Supplementary Information section S-3. While there is no generally accepted meaning of the phrase ``fully variational'', and it is not clear exactly in what sense the authors use this phrase, the authors repeatedly imply that the CBE algorithm necessarily lowers the energy of the MPS with each step. It is important to be precise in the language on this point, because if this was really true then it
would put the CBE algorithm within the class of algorithms known as Hill Climbing\cite{Hill}, named since every iteration of the algorithm, if the energy changes at all then it necessarily moves up hill (lower energy). Pure single-site DMRG is an example of a hill climbing algorithm. Such algorithms are notorious for slow convergence, since if the system gets into a local minima then it cannot temporarily increase the energy to escape. This explains why single-site DMRG is notoriously slow to converge, and if good quantum numbers are used then it will usually never reach the variational minimum unless the starting state is already very close, because there is no way for the algorithm to adapt the relative sizes of each quantum number sector; every choice of quantum number distribution leads to a different local minima. Fortunately, the CBE algorithm, similarly to 2-site DMRG, single-site mixing\cite{White1s}, 3S\cite{3S}, AMEn\cite{DolgivAMEn} etc., is \emph{not} a hill climbing algorithm, due to the truncation step (CBE update step (iv)). Just like other bond expansion schemes, the singular value decomposition at the end of the optimization phase of each step truncates the bond dimension from a larger space back to $D$ states. In 2-site DMRG and density matrix mixing (including 3S), this is a truncation from up to $dD$
non-zero singular values down to $D$. In CBE, it is a truncation from $D + k = D + \tilde{D}$ to $D$. Because of the truncation at each step, these algorithms never converge to a unique fixed point: the bond expansion breaks translation symmetry and the energy of the wavefunction is different at each active site of the lattice. The best that can be achieved with a fixed number of expansion vectors is a limit cycle, where the wavefunction returns to the same point after two complete half-sweeps;  the kept and discarded states are slightly different at every step and the 2-site discarded space $\disc\disc$ is never empty. Thus the entirety of Ref.~\cite{CBE} endnote [29] is based on a false premise, and to the contrary the expansion vectors added to the basis drive the calculation away from the optimal fixed point of a bond dimension $D$ MPS. This is readily seen by examining the energy as a function of position in the lattice during the DMRG sweeping, as the bond expansion causes a shallow bound state across the expanded bond\cite{TwoSite,TwoSiteComment}, with a bond energy that can even be below that of the ground state\footnote{This does not violate the variational principle, because only the \emph{total} energy of the entire system is variational. If one bond has an energy below that of the ground state it will be compensated by higher bond energies elsewhere. This is why DMRG with bond expansion never reaches the exact variational minimum for fixed bond dimension\cite{Variational}.}. A close examination of the figures in Ref.~\cite{CBE} shows that the energy is not monotonically decreasing, although it is difficult to see as the variation is smaller than the line widths of the plots. So while the upwards fluctuations in energy in CBE might be much smaller than in 2-site DMRG, the fact that they can exist is very important for the convergence properties of the algorithm.

\begin{figure*}[t]
   \includegraphics[width=\textwidth]{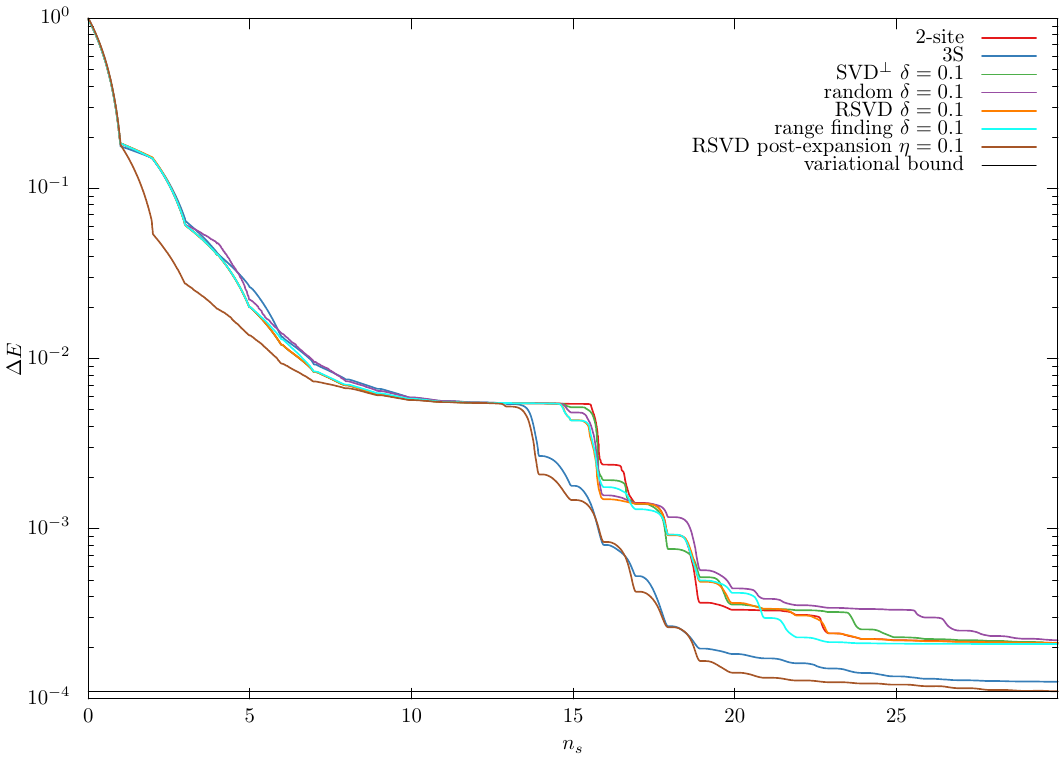}
   \caption{Convergence in the relative error for non-interacting fermions with PBC, otherwise same parameters as used in
   Ref.~\protect{\cite{CBE}} Fig.~S-6 with bond dimension $D=600$. 3S result uses a mixing parameter of $\alpha=10^{-4}$. $\text{SVD}^\perp$ is a baseline comparison against CBE, using a full SVD of the tensor $X_{\bar{A}}$ projected onto the 2-site tangent space. \emph{Random} uses $k = \delta D$ Haar-distributed random vectors. \emph{RSVD} and \emph{range finding} are the $O(dwkD^2)$ algorithms discussed in the main text. \emph{RSVD post-expansion} uses a new variant of 3S where $k=\eta D$ expansion vectors are added to the MPS after the truncation step. The \emph{variational bound} is the best energy we could obtain for fixed bond dimension $D=600$. In the first half of the calculation, the 2-site and $\text{SVD}^\perp$ lines are hidden by that of RSVD, which in turn is mainly hidden by the range finding curve.}
   \label{fig:f1}
\end{figure*}

\noindent \textbf{Comparison to 3S --} The idea behind the single-site density matrix mixing scheme\cite{White1s} and the 3S algorithm\cite{3S} is to incorporate degrees of freedom into the kept states that are likely to be useful to represent terms in the Hamiltonian, even if they do not (yet) do anything at the active site. By adding such states to the basis throughout a sweep, matrix elements corresponding to long range interactions can be incorporated even if they never appear at any order in perturbation of 2-site projected Hamiltonians. Perhaps the clearest example is models with an inhomogeneous unit cell, such as a lattice gauge theory where gauge sites do not contain any matter degrees of freedom, and hence a 2-site update that consists of one matter and one gauge site cannot incorporate new matter quantum numbers. The mixing term overcomes this problem, but the cost is that some proportion of the kept states are ultimately not so useful with respect to the optimal ground state. The magnitude of the perturbations is controlled by the \emph{mixing parameter} $\alpha$, and although one could try to gradually reduce $\alpha$ all the way to zero throughout the calculation, this is not an efficient process and there is still no guarantee that this leads to the variational minima.
It is a generic feature of fixed bond dimension MPS calculations that converging much beyond the leading digit of the relative error is an inefficient use of computational resources anyway, and if you want to improve the relative error then it is much better to simply increase the bond dimension a few percent. That is, the best performance is obtained by \emph{continuing to increase the bond dimension}, only stopping once the desired error is achieved. For most purposes the resulting wavefunction is good enough, but if a specific calculation requires something closer to the variational bound for a lower bond dimension, then it is often better to approach the variational energy from \emph{below}, starting from a larger bond dimension and progressively reducing it, perhaps combined with a small mixing term (or bond expansion) that is smoothly taken to zero at the same time. The authors own comparisons of CBE versus 3S in figures S-7 and S-8 demonstrate this: the most efficient way to get to a given relative error is to continue increasing the bond dimension until the desired energy error is achieved. It is in this sense that 3S is much more efficient than 2-site DMRG\cite{3S}. It is surely true that CBE is more efficient than 3S in systems with short-range interactions, but the authors own calculation in figure S-7 shows that the difference is small: although there is a long tail for 3S to converge for a fixed bond dimension, this is entirely expected and the important message of figure S-7 is that the most efficient way to lower the variational energy is to continue increasing the bond dimension. Indeed, the figure shows that this only takes \emph{one additional sweep} with increased bond dimension, with computational resources that are only slightly higher than CBE. We note that the authors recommend using 3S-style mixing in the early stages of a calculation anyway\footnote{Section S-3(B), final paragraph}, but in some cases this isn't enough and the mixing is required throughout.

None of the example calculations presented in Ref.~\cite{CBE} contain interactions beyond 6 sites. It has been understood for a long time\cite{White1s,3S} that 2-site DMRG (and hence also CBE and other algorithms based on expanding the environment bond prior to the optimization step; we refer to this class of algorithm as \emph{local pre-expansion}) can have convergence problems with long range interactions. To illustrate this, we take an example from Ref.~\cite{CBE}, for spinful non-interacting fermions on an $L=100$ site chain but now making the boundary condition \emph{periodic}, and pre-expansion algorithms struggle to converge. This is shown in \reffig{fig:f1}, where we compare several variants of pre-expansion including 2-site DMRG and the randomized SVD approach described above, and density matrix mixing via the 3S algorithm\cite{3S}. All of the algorithms get temporarily stuck in a metastable excited state, which corresponds to the \emph{open} boundary condition wavefunction. It takes some time for the basis to grow sufficiently to contain Hamiltonian matrix elements that connect the periodic boundary, and thereafter the convergence resumes. But even after overcoming the metastable state, local bond expansion algorithms have trouble, often encountering additional metastable plateaus, and none of the local expansion algorithms get the relative error closer than a factor $\sim 2$ of the variational bound for $D=600$ after 30 sweeps. However it is notable that all of the presented algorithms, including random expansion vectors, converge very well in the initial steps, indicating that in the absence of long range interactions the choice of algorithm is not a decisive factor. This is not a detail of our particular implementation of DMRG, and we verified that the behavior of the 2-site DMRG code in the iTensor toolkit\cite{itensor} is consistent with these results\footnote{As iTensor does not support non-Abelian symmetries, we used a $U(1)\times U(1)$ basis for the iTensor calculations. This is not directly comparable to the $U(1)\times SU(2)$
results shown here, but the qualitative behavior is the same.}. The algorithms escape from the metastable state rather chaotically, and this depends very sensitively on the numerical details. For example, decreasing the $\epsilon$ tolerance of the eigensolver might help the calculation to escape the metastability, or it might cause the calculation to fall even deeper into the metastable state. We don't know of any systematic way to tune these parameters in this situation, so we simply used the recommended default parameters used by the Matrix Product Toolkit\cite{Toolkit}. The \emph{random}, \emph{RSVD}, and \emph{range finding} algorithms use randomized sketching and are therefore non-deterministic; for these cases we ran 10 instances of the same calculation and chose a sample that best represents the median convergence. The separate runs produce near-identical results for the first half of the calculation, and only differ once they escape the metastable state. However the distribution of the results for each algorithm is quite distinct such that the depiction in \reffig{fig:f1} is representative of the expected relative performance.  \reffig{fig:f1} also includes a calculation using \emph{post-expansion}, which is a term we use to describe the class of algorithms that expand the basis after the optimisation step. This class includes the 3S algorithm as well as a new algorithm that uses the randomized SVD of the 3S mixing term to construct $k$ expansion vectors in the null space of $C$, which are added basis after the truncation giving an MPS with a bond dimension of $D+k$. This can also be done in computation time $O(dwkD^2)$, and has an appealing physical interpretation. Full details of these algorithms will be published separately\cite{Forthcoming}.

\begin{acknowledgments}
I.P.M.~acknowledges funding from the National Science and Technology Council (NSTC) Grant No.~122-2811-M-007-044. The calculations presented here were obtained using the Matrix Product Toolkit\cite{Toolkit}.
\end{acknowledgments}

\appendix

\section{Appendix A: Additional details and benchmarks}

\begin{figure}[ht]
   \includegraphics[width=\columnwidth]{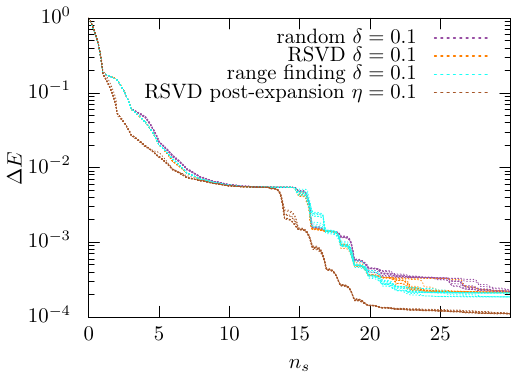}
   \caption{All runs of the randomized algorithms from \protect{\reffig{fig:f1}}.}
   \label{fig:f1detail}
\end{figure}

In \reffig{fig:f1}, four of the algorithms are based on randomized sketching (\textit{random}, \textit{RSVD}, \textit{range finding}, \textit{RSVD post-expansion}). While the performance of these algorithms is generally not random, the escape from a metastable state is rather chaotic and small changes can affect the calculation, even for notionally deterministic algorithms. To obtain \reffig{fig:f1} for the randomized algorithms, we ran each calculation 10 times and selected one run that we thought best represented the median performance. We show all of the runs in \reffig{fig:f1detail}. Four out of the ten runs using the range finding algorithm actually found a lower energy state than 2-site DMRG, showing that some randomness can help escape metastability, even though the 2-site algorithm is exploring a larger local Hilbert space.

\begin{figure}[ht]
   \includegraphics[width=\columnwidth]{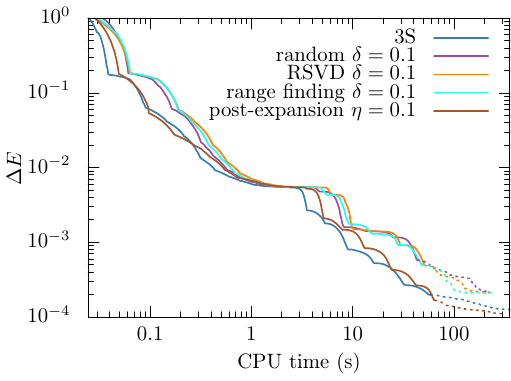}
   \caption{Log-log plot of the relative error in the energy versus CPU time in seconds for the $3S$, pre-expansion and post-expansion algorithms from \protect{\reffig{fig:f1}}. Solid lines are the bond expansion phase, as the bond dimension $D$ is increased to $600$. Dashed lines are a further 11 sweeps with $D=600$.}
   \label{fig:cpu}
\end{figure}

\reffig{fig:cpu} shows the CPU time for the $3S$, pre-expansion and post-expansion algorithms. Calculations were performed on a single core Intel Core i7-12700KF CPU\footnote{According to the GeekBench 6 benchmark at \url{https://browser.geekbench.com} this processor has approximately 89\% faster single-core performance than the i7-9750H CPU used in Ref.~\cite{CBE}.}, with double precision complex arithmetic\footnote{Although though the Hamiltonian is purely real, our code currently only allows complex arithmetic, and we used complex Gaussian random matrices. In principle, real arithmetic would be $\sim$ 3-4 times faster}. With the local Hilbert space dimension of 3 multiplets, the $O(d)$ vs $O(d^2)$ performance improvement of the new algorithms is not large. We argue in the main text that the most relevant performance comparison is during the bond growth phase of the calculation, which is indicated in solid lines, transitioning to dashed lines for the final part of the calculation with fixed $D$.

\section{Appendix B: Hubbard-Holstein model}

\begin{figure}[ht]
   \includegraphics[width=\columnwidth]{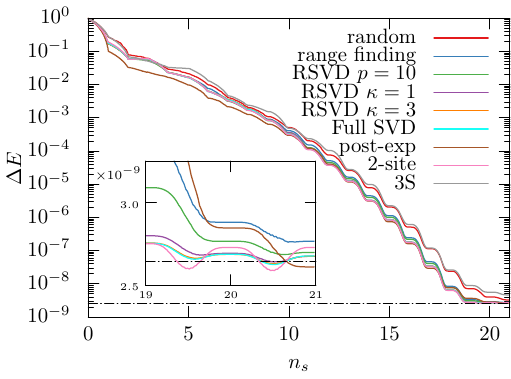}
   \caption{Convergence of the Hubbard-Holstein model, energy versus number of half-sweeps. The random, range finding, and RSVD algorithms all use $\delta=0.1$ bond expansion. Post-exp is post-expansion using RSVD with $p=10$ oversampling. The inset shows the detailed convergence over the last two sweeps where the bond dimension is fixed at $D=600$. The black dashed line is an estimated variational bound for $D=600$. Post-expansion ads $\eta=0.1$ expansion states after the truncation phase, leading to an MPS with a final bond dimension of $D(1+\eta)=660$.}
   \label{fig:hh}
\end{figure}

As a second benchmark, we examine the Hubbard-Holstein model, using the same parameters as Fig.~3 of Ref.~\cite{CBE}, with maximum phonon occupancy of 3. Focusing on bond dimension $D=600$, \reffig{fig:hh} shows the relative energy error per sweep for a selection of different local pre-expansion algorithms. \emph{Full SVD} is a baseline comparison using the full SVD of the expansion tensor $X_{\bar{A}}$ (note we did \emph{not} project out the kept states of the $C$ tensor). \emph{Random} vectors do not work so well in this model, probably because of the large local Hilbert space dimension of 12 states means that the $k=0.1D = 60$ expansion states out of a possible 660 has a relatively low probability of making a good selection. The \emph{range finding} algorithm works very well though. Note that for the random and range finding algorithms, we force a minimum of 1 state per quantum number sector into the expansion basis. Without sampling all of the available symmetry sectors, the random and range finding algorithms are ineffective. We also show three settings for oversampling of the RSVD; the baseline $p=10$ oversampling suggested in Ref.~\cite{RSVD}, and multiplicative oversampling with $p = \min(k+10,\kappa k)$, for $\kappa = 1$ and $\kappa=3$.
Multiplicative oversampling has a theoretical truncation error of a factor $\simeq 1 + 1/\kappa$ larger than the exact SVD\cite{RSVD}, and $\kappa=1$ is more than sufficient for this calculation. We also show 3S and 2-site DMRG results for comparison. Aside from the \emph{random} and \emph{3S} curves, the remaining curves are basically on top of each other.
The inset shows the detail of the convergence in the final sweeps. For pre-expansion algorithms, the energy can drop below the variational bound for $D=600$ because of the larger Hilbert space during the bond expansion, which leads to a shallow bound state and a lower energy at that bond\protect{\cite{TwoSite,TwoSiteComment}}. At the end of the sweep, when all of the MPS bonds are dimension $\leq 600$, the energy is above the estimated variational bound. Because the RSVD post-expansion algorithm results in an MPS with an enlarged bond dimension $D(1+\eta)$, it is allowed that the final energy is below the estimated variational bound and this actually occurs at the end of the calculation.

\begin{figure}[ht]
   \includegraphics[width=\columnwidth]{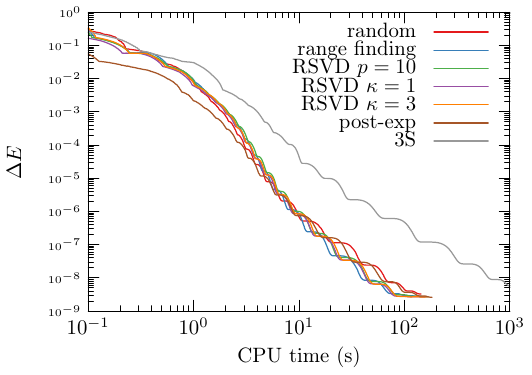}
   \caption{CPU time (seconds) for the Hubbard-Holstein model.}
   \label{fig:hh-cpu}
\end{figure}

For completeness, we show the CPU time for the Hubbard-Holstein calculation in \reffig{fig:hh-cpu}. All of the pre-expansion and post-expansion algorithms perform similarly. $3S$ suffers from the larger local Hilbert space due to the $O(d^2)$ scaling. Random pre-expansion has a somewhat worse energy, and post-expansion is somewhat slower because both sides of the active tensor have an expanded bond dimension, i.e., in pre-expansion on a right-to-left sweep the $C$ tensor has dimension $(D+k)\times d \times D$, but for post-expansion it has dimension $(D+k)\times d \times (D+k)$. However, post-expansion has a significant advantage, in that it works well even when there are long-range interactions. For the Hubbard-Holstein model this means that it is possible to factorize the unit cell of the lattice into separate fermion and phonon sites, which enables a significant increase the number of phonon modes. With $d_{\text{ph}}$ phonon modes (corresponding to a maximum phonon occupancy of $d_{\text{ph}}-1$), the coarse-grained lattice site has a local Hilbert space of $3 d_{\text{ph}}$ multiplets ($4 d_{\text{ph}}$ without using $SU(2)$ symmetry for the fermions), whereas splitting the unit cell into two lattice sites gives one site for 3 fermion states and one site for the
$d_{\text{ph}}$ phonon modes. In principle the CPU time is then proportional to $3+d_{\text{ph}}$, rather than $3 d_{\text{ph}}$. To tackle even larger values of $d_{\text{ph}}$, the phonon modes can themselves be factorized into multiple sites\cite{Jeckelmann}. Unfortunately, this procedure fails completely for local pre-expansion algorithms, including 2-site DMRG, because on every pair of neighboring lattice sites there is only at most one fermion site, and hence expanding the bond cannot introduce new fermionic quantum numbers.

\begin{figure}[ht]
   \includegraphics[width=\columnwidth]{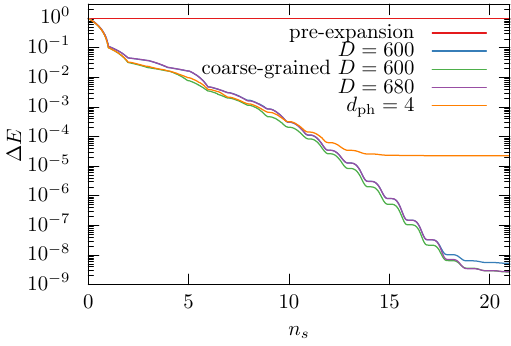}
   \caption{Convergence of the Hubbard-Holstein model increasing the maximum phonon number to $8$ ($d_{\text{ph}}=9$ phonon modes). For comparison purposes, the $D=600$ result for 3 phonons ($d_{\text{ph}}=4$) is included showing that the limiting factor affecting the ground state energy is the number of phonon modes. The coarse-grained curve uses the basis of the combined fermion+phonon. All pre-expansion algorithms fail completely and the energy stays at 0. }
   \label{fig:hh8}
\end{figure}
On increasing the maximum number of phonons, the advantage of using a fine-grained unit cell becomes very clear. \reffig{fig:hh8} shows the convergence for the same calculation as \reffig{fig:hh} but now with up to 8 phonons per site ($d_{\text{ph}}=9$). Using a unit cell of two sites, there are now a total of 100 sites in the lattice (50 fermion sites and 50 phonon sites), and post-expansion works extremely well, producing an energy that is nearly as good as for the coarse-grained lattice. Splitting the fermion and phonon sites slightly restricts the entanglement allowed between them compared with the coarse-grained lattice, which affects the energy slightly. This can be compensated by increasing the bond dimension, and $D=680$ is sufficient here. Pre-expansion algorithms, including 2-site DMRG, fail completely with the factorized basis.

\begin{figure}[ht]
   \includegraphics[width=\columnwidth]{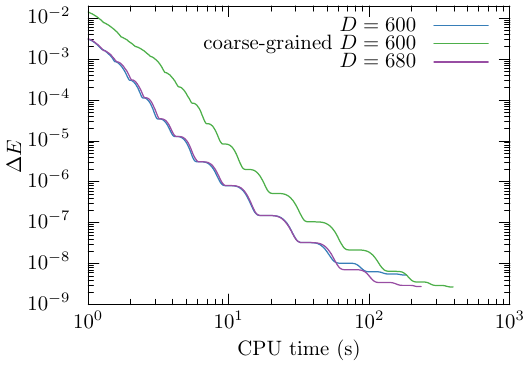}
   \caption{CPU time for the calculation of \protect{\reffig{fig:hh8}}. }
   \label{fig:hh8-cpu}
\end{figure}

The CPU time for this calculation is shown in \reffig{fig:hh8-cpu}, which shows the expected $\simeq 2$ speedup from using a factorized basis, going from local Hilbert space sizes of $3 \times 9 = 27$ to $3 + 9 = 12$. This advantage clearly gets bigger for even larger phonon numbers needed for the Holstein model when the zero mode is not removed from the basis\cite{Jeckelmann,HolsteinBasis}.

\section{Appendix C: Random sketching versus shrewd selection}


\begin{figure}[ht]
   \includegraphics[width=\columnwidth]{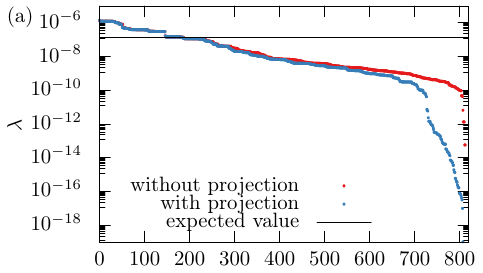}\par
   \includegraphics[width=\columnwidth]{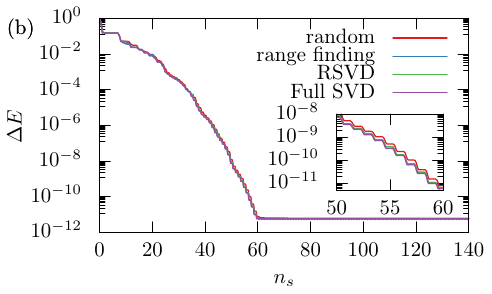}
   \caption{Comparison using randomized sketching for the expansion vectors using the same parameters as figure S-5 of Ref.~\protect{\cite{CBE}}.
   (a) the singular values of the expansion matrix $X$ without projecting onto the two-site tangent space and $X_{\bar{A}}$ with the projection.
   (b) difference in the variational energy from the exact solution, starting from a $D=1$ valence bond state increasing to $D=300$. The curve for the range finding algorithm is almost indistinguishable from (and hidden by) the RSVD curve. For the RSVD, we used oversampling factor $\kappa=1$, with a minimum oversample of 10.
   }
   \label{fig:SingularComparison}
\end{figure}

In this section, we compare in detail the effectiveness of the algorithms based on random sketching. In \reffig{fig:SingularComparison} (a), we compare the singular values of the expansion tensor $X$ with and without the projection onto the 2-site tangent space, for bond dimension $D=300$ at the center of the $L=20$ site chain. The projection makes very little difference to the top part of the spectrum, except to give very low weight to some states that would otherwise appear around the middle of the spectrum. The bulk of the spectrum has a very steady exponential decay, which explains why random expansion vectors work so well: a Haar random vector has an expected weight of nearly 10\% of the leading singular value. \reffig{fig:SingularComparison} (b) shows the convergence of the energy for the different algorithms. The full SVD is included as a baseline. As expected from the unchanged top part of the spectrum in \reffig{fig:SingularComparison} (a), it makes no difference in this case whether the expansion tensor is projected onto the two-site tangent space or not.
As best as we can tell, random vectors are already very close to the `moderate pre-selection' line of figure S-5 of Ref.~\cite{CBE}, and range-finding exeeds it. However random vectors are much less effective when only a few states from the local Hibert space are used, such as the Hubbard-Holstein model. In that case, the phonon number distribution is far from uniform, and only a small subset of the null space are effective expansion vectors and we recommend that at least range-finding is used, and preferably the RSVD. When using good quantum numbers, there is a question of how to choose the quantum number sectors of the expanded states. Our approach is to choose sectors such that the distribution of quantum numbers matches the the states already kept, and in addition we require at least one expansion vector in each available quantum number sector. This typically results in slightly more than $\delta D$ expansion vectors, but the cost of this is negligible as the computation time is dominated by the matrix multiplies in the largest quantum number sectors. For the post-expansion algorithm, this requirement to explore all candidate quantum number sectors means that random vectors or random range-finding algorithms are not so effective, since there is no singular value truncation to control the number of quantum number sectors, which can therefore get very large. For this reason, we recommend using the RSVD with post-expansion, since the singular value selection serves to prevent the number of quantum number sectors from growing out of control. But even without quantum number symmetries, it is best to use the RSVD with post expansion. This is because in post-expansion the vectors are used for the entire sweep so it is more important that the vectors are chosen as to have a high probability of contributing to the converged wavefunction. In contrast, in the pre-expansion algorithm the vectors are selected from the null space of the environment tensor and they are discarded at the end of the current iteration, so a single iteration with an unlucky or poor selection of expansion vectors is of no major consequence, as long as there is a good average selection throughout multiple sweeps.

\bibliography{main}

\end{document}